\documentclass[preprint,12pt]{elsarticle}
\usepackage{graphicx} 
\usepackage{amsmath}
\usepackage{url}
\usepackage{booktabs}
\usepackage[table]{xcolor}
\definecolor{mypurple}{RGB}{120,80,180}
\definecolor{mypink}{RGB}{230,80,180}
\definecolor{cedar}{RGB}{200,215,150}
\definecolor{mygreen}{RGB}{94,170,55}
\definecolor{myblue}{RGB}{140,200,255}

\usepackage{subcaption}
\usepackage{amssymb}
\usepackage{booktabs}
\usepackage{graphicx}
\usepackage{subcaption}
\usepackage{bm}        

\journal{Biomedical Informatics}

\begin{document}
\begin{frontmatter}




\title{A Correspondence Score for Validating Learned Patient Representations in Clinical Tabular Data}


\author{Majid Lotfian Delouee, Sjors G. J. G. In ’t Veld, Martijn C. Schut} 

\affiliation{organization={Translational AI Laboratory, Department of Laboratory Medicine, Amsterdam UMC},
            country={The Netherlands}}





\begin{abstract}

\noindent
\textbf{Objective:}
Identifying patients with similar clinical characteristics is important for cohort discovery, clinical decision support, and personalized medicine. Recent AI methods increasingly summarize electronic health record data into learned patient representations to support these applications. However, it is often unclear whether these representations preserve the similarities observed in the original clinical data. Consequently, patients who appear similar in a learned representation may not reflect the same similarities in the original data, potentially affecting subsequent clinical and research applications. Existing evaluation approaches rely primarily on task-specific performance metrics and do not directly assess how well patient similarity is preserved.
Therefore, we propose a correspondence score to assess whether learned patient representations preserve clinically meaningful relationships observed in the original data. With this score, we investigate how correspondence varies across representation learning methods and representation dimensionalities.

\noindent
\textbf{Methods:}
We introduce a correspondence score that combines multiple metrics which are measuring the preservation of data relationships, with weights estimated through Bayesian optimization to account for dataset-specific structural properties. 
The approach was evaluated using routine complete blood count (CBC) data, a common clinical test that captures patient characteristics relevant to conditions such as anemia and inflammation. Representations generated by five tabular foundation models and six classical transformation techniques were assessed across four datasets to evaluate the robustness and consistency of correspondence estimates across different patient populations and data distributions.

\noindent
\textbf{Results:}
The correspondence score revealed substantial differences in how well learned representations preserve patient relationships across models and representation dimensionalities.
Patient similarity was best preserved at moderate representation dimensionalities, whereas both very low- and very high-dimensional representations introduced measurable distortions. Across the evaluated settings, correspondence was generally highest at intermediate dimensionalities, while more aggressive reduction or expansion was associated with greater deviation from the original data structure.
Foundation models preserved correspondence more consistently than classical approaches, with improvements ranging from approximately 5\% to over 125\%, providing principled, pre-deployment justification for method selection.
These findings demonstrate that the proposed score enables actionable validation of learned representations, supporting informed decisions about model choice and dimensionality independently of labeled outcomes, before representations are committed to clinical use.

\end{abstract}


\end{frontmatter}


\section{Introduction}
\label{intro}

Patient similarity is fundamental to many biomedical informatics applications, including cohort discovery, disease subtyping, risk stratification, and clinical decision support. These applications rely on identifying patients who share similar clinical characteristics and grouping them into meaningful populations for analysis and decision making \cite{Murali2023EHRKG, Amirahmadi2023EHRTrajReview}. As healthcare data become increasingly available through electronic health records and routine clinical testing, accurately characterizing similarities among patients has become an important challenge in biomedical informatics \cite{Kauffman2025EHREmbeddings}.

A central feature of modern machine learning models is that they do not simply process clinical data; they transform it \cite{Xian2025TransformerPatientEmbedding}. During training, representation-learning models convert tabular patient records into compact numerical summaries, commonly called embeddings or learned representations, that capture patterns across many clinical variables simultaneously. Unlike classical approaches such as logistic regression or gradient boosting, which operate directly on the original features, these models produce reusable intermediate representations that can be extracted and applied across multiple analyses. Recent developments in tabular foundation models have accelerated this trend, with representations now being shared and reused across datasets, institutions, and downstream tasks \cite{Timilsina2025HarmonizingFMs, Hollmann2025TabPFN, moor2023foundation}.

A key consideration when reusing learned representations is that their structure is influenced by the objective used during training. For example, a model trained to predict 30-day hospital readmission will organize patients according to patterns that are informative for readmission risk. As a result, the learned representation may capture some aspects of the original clinical data more strongly than others. Patient characteristics that are highly relevant to the training objective may become more prominent, while other clinically meaningful distinctions, such as differences between anemia subtypes or inflammatory profiles, may be represented less clearly. This is not a limitation of any particular modeling approach; rather, it is a natural consequence of learning representations optimized for a specific objective. Consequently, the structure of a learned representation should not be assumed to fully reflect the structure of the original clinical data, particularly when it is reused for purposes beyond the task on which it was trained.

\begin{figure}[!htb]
    \centering
    \includegraphics[width=\linewidth , trim=0cm 4cm 7cm 2cm, clip]{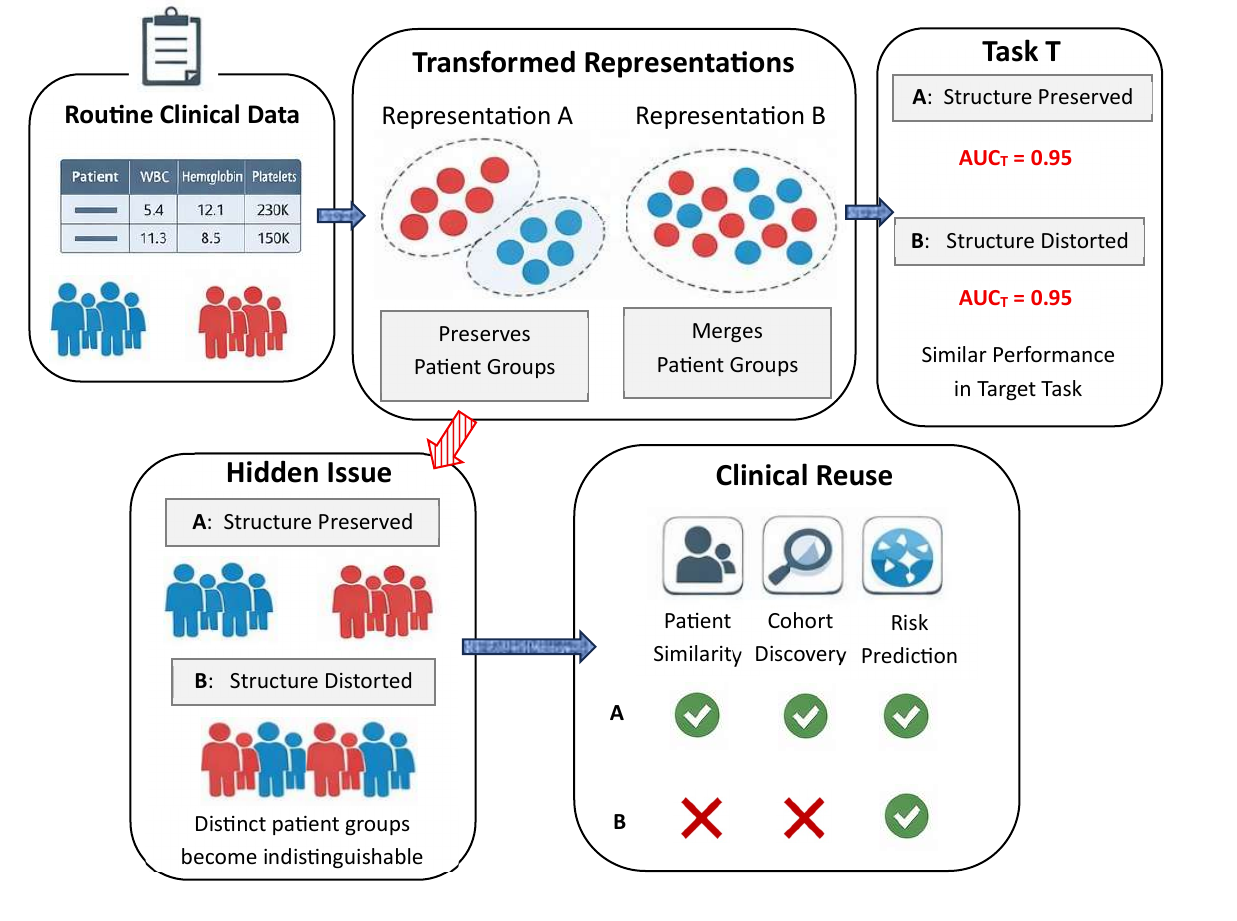}
    \caption{Two representations derived from the same routine clinical data achieve identical performance on a single downstream task T (AUC = 0.95), yet differ substantially in how they organize patient relationships. Representation A preserves the structure of the original data, keeping clinically distinct patient groups separated; Representation B merges them into a single undifferentiated cluster. Because task evaluation scores both representations equally, this difference is invisible at the point where model selection decisions are typically made. The consequence only becomes apparent when the representations are reused: Representation A supports patient similarity search and cohort discovery as well as risk prediction, while Representation B fails on the two tasks that depend on preserved patient neighborhoods. The correspondence score introduced in this study measures this gap directly, before representations are committed to downstream use.}
    \label{fig:intro}
\end{figure}

When such representations are reused for tasks they were not trained on (e.g., patient similarity search, cohort stratification, unsupervised phenotyping) the distortions travel with them. As illustrated in Figure~\ref{fig:intro}, two representations of the same dataset may achieve comparable predictive performance while organizing patients in substantially different ways. One may keep patients with microcytic anemia clearly separated from those with inflammatory anemia; another may merge those groups entirely. These differences are invisible to task-specific performance metrics, yet they directly determine the clinical conclusions drawn from similarity-based retrieval, clustering, and subgroup analysis. Without a way to assess whether a representation corresponds to the observed patient data, such distortions remain undetected until they have already influenced a clinical interpretation.

The evaluation of learned representations is broadly divided into extrinsic and intrinsic approaches. Extrinsic evaluation measures how well representations support a specific downstream task, such as classification or survival prediction. It is the dominant paradigm in biomedical machine learning, but it answers only one question: how useful is this representation for this task? It cannot answer whether the representation faithfully reflects the relationships present in the observed data. Intrinsic evaluation attempts to fill this role by examining how well the structure of the original data is preserved after transformation, using measures such as trustworthiness, continuity, and rank-based distance correlation \cite{venna2001neighborhood, lee2009quality}. These measures exist and are well understood individually, but they are almost always reported in isolation. No unified, interpretable correspondence score currently exists that combines them into a practical validation instrument for clinical AI workflows.

This gap matters because it leaves a critical step missing from the clinical AI development pipeline. Before a learned representation is deployed for patient similarity search, used to construct a clinical cohort, or shared across institutions, there is currently no standard way to verify that it preserves the patient relationships present in the observed data. This study addresses that gap directly. We introduce a correspondence score that quantifies how well learned representations correspond to the observed clinical data, without requiring downstream task labels. The score combines multiple complementary structure-aware measures into a single interpretable value through dataset-adaptive Bayesian optimization, and it is designed to work in both dimensionality reduction and expansion settings, the two directions in which representations are most commonly generated in biomedical informatics.

\begin{table}[!t]
\renewcommand{\arraystretch}{1.3}
\setlength{\tabcolsep}{6pt}
\centering
\begin{tabular}{p{0.18\linewidth} p{0.82\linewidth}}
\toprule
\multicolumn{2}{c}{\textbf{Statement of Significance}} \\[2pt]
\midrule

\textbf{\small Problem or Issue:} &
Learned patient representations are increasingly reused for applications such as patient similarity search, cohort construction, and phenotyping. Because the structure of these representations is influenced by the objective used during training, different representations may preserve different aspects of the original clinical data. Existing evaluation approaches focus primarily on downstream task performance and provide limited insight into how well learned representations reflect the underlying patient structure, creating a validation gap in clinical AI development. \\[6pt]

\textbf{\small What is Already Known:} &
Intrinsic measures such as trustworthiness, continuity, and rank-based distance metrics can assess how well representations preserve relationships present in the original data. However, these measures are typically applied in isolation and do not provide a unified, interpretable assessment of representation correspondence suitable for routine validation in biomedical informatics workflows. \\[6pt]

\textbf{\small What this Paper Adds:} &
A correspondence score that fills the missing validation step for representation reuse in clinical AI. By combining complementary similarity-preservation measures through dataset-adaptive Bayesian optimization, the score enables principled assessment of how well learned representations reflect the original patient data before they are reused in downstream analyses. \\[6pt]

\textbf{\small Who Would Benefit:} &
Clinical data scientists, biomedical informaticians, and research teams who develop, evaluate, or reuse learned patient representations and require a principled validation approach that is independent of downstream task performance. \\

\bottomrule
\end{tabular}
\end{table}

We applied the proposed correspondence score to representations generated from four real-world clinical datasets based on routine blood count measurements, comparing five tabular foundation models against six classical transformation techniques across a range of representation dimensionalities. The results reveal that correspondence does not improve monotonically with model complexity or dimensionality. Instead, it peaks at intermediate sizes, while in both low- and high-dimensionality the preservation of patient relationships is degraded. Foundation models preserved correspondence more consistently than classical approaches, with improvements ranging from 5\% to over 125\%, and representations with similar predictive performance were shown to differ substantially in correspondence, a difference that task-based evaluation alone would never detect.

Our main contributions are as follows:

\begin{itemize}

    \item \textbf{A correspondence score as a missing validation step:}
    We introduce a principled, task-independent score that quantifies how well learned clinical representations correspond to the observed patient data, providing direction-specific values for both low- and high-dimensionality settings.

    \item \textbf{Validation on real clinical data:}
    We apply the score to representations from four real-world blood count datasets, enabling systematic comparison across methods, model families, and representation dimensionalities.

    \item \textbf{Actionable guidance for representation design:}
    We demonstrate that learned representations can differ substantially in correspondence across models and representation dimensionalities, and that both model choice and dimensionality have systematic, detectable effects, findings that directly support pre-deployment validation decisions.

\end{itemize}

\section{Related Work}
\label{sec:relatedwork}

The evaluation of learned representations sits at the intersection of machine learning methodology and clinical data science. Prior work in this space has largely focused on how well representations support specific predictive tasks, while the question of whether they preserve the patient relationships present in the observed data has received comparatively little systematic attention \cite{Amirahmadi2023EHRTrajReview, Zheng2025SelfSupervisedEHR}. We review the lines of work most relevant to our contribution: the extrinsic and intrinsic evaluation of representations, dimensionality transformation methods and their structural effects, tabular foundation models and how their representations are assessed, efforts to combine intrinsic measures into unified scores, and the role of learned representations in clinical workflows.

\subsection{Extrinsic and Intrinsic Evaluation of Representations}

The standard approach to evaluate a learned representation is to measure how well it supports a downstream task, predicting an outcome, classifying a condition, or stratifying a risk group. This extrinsic paradigm dominates biomedical machine learning for practical reasons: labeled outcomes are the ultimate test of clinical utility, and task performance is straightforward to compute and compare \cite{bommasani2021opportunities, bengio2013representation}. However, it answers only a narrow question. Strong performance on a specific task reflects how well the representation captures task-relevant structure; it says nothing about whether the representation preserves the broader pattern of relationships present in the observed patient data.

Intrinsic evaluation offers a complementary perspective by examining the correspondence between a learned representation and the observed data, without relying on any outcome label \cite{venna2001neighborhood}. Established intrinsic measures include trustworthiness and continuity, which assess how well local neighborhoods are preserved \cite{venna2010information}; Shepard correlation, which captures the rank-order consistency of pairwise distances across the original and transformed spaces \cite{lee2009quality}; and kNN preservation and pairwise distance error, which quantify local and global distortion respectively \cite{espadoto2019toward}. Each of these metrics illuminates a specific geometric property of the representation, and each has a clear clinical interpretation: trustworthiness asks whether patients that appear close in the representation were actually close in the observed data, while continuity asks the reverse.

In clinical settings, the limitations of purely extrinsic evaluation become especially consequential. Outcome labels are often delayed, incomplete, or aligned with only one of the many purposes for which a representation might be reused \cite{Kauffman2025EHREmbeddings}. A representation evaluated solely on its readmission prediction performance provides no assurance of its suitability for cohort construction or phenotype discovery. Intrinsic evaluation addresses this by asking whether the structure of the observed data survives the transformation, a question that is clinically relevant whenever representations are reused beyond their original task.

\subsection{Dimensionality Transformation and Its Structural Effects}

Clinical tabular data is routinely transformed to manage its complexity. Dimensionality reduction methods such as PCA, UMAP, and variational autoencoders project patient data into lower-dimensional spaces that are easier to visualize and analyze \cite{becht2019dimensionality, diaz2019umap}. Dimensionality expansion methods, including polynomial feature mappings, random projections, and decoder-based mappings, increase representational capacity by introducing additional dimensions that may capture nonlinear interactions \cite{DudaHart2nd, bingham2001random, rahimi2007random, kingma2013auto}.

Both directions of transformation change the geometry of the data, and both can alter how patients relate to one another. A reduction that compresses too aggressively may merge patients from distinct clinical groups; an expansion that introduces redundant dimensions may dilute meaningful neighborhood structure. Despite these effects, the standard practice for evaluating such methods remains downstream task performance or visual inspection, neither of which provides a direct, quantitative measure of how much relational structure has been preserved or lost. Our work is motivated in part by this gap: the structural effects of dimensionality transformation on patient relationships deserve explicit, systematic assessment.

\subsection{Representation Learning with Tabular Foundation Models}

Foundation models have reshaped representation learning across language \cite{devlin2018bert, radford2019language}, vision, and increasingly tabular data \cite{bommasani2021opportunities}. In the clinical domain, architectures such as FT-Transformer \cite{gorishniy2021fttransformer}, TabNet \cite{arik2021tabnet}, SAINT \cite{somepalli2021saint}, TabTransformer \cite{huang2020tabtransformer}, and TransTab \cite{wang2022transtab} have been developed to handle the particular challenges of heterogeneous structured data, mixed feature types, missingness, and variable coding practices. These models learn rich internal representations during training, and those representations are increasingly extracted and reused for analyses beyond the original training task \cite{Timilsina2025HarmonizingFMs}.

Surveys of tabular foundation models in healthcare report strong predictive performance across a range of clinical tasks \cite{zhou2023comprehensive, moor2023foundation, schmidt2023recent}. However, existing evaluations almost exclusively rely on extrinsic metrics such as AUROC or accuracy. Whether these models produce representations that correspond to the structure of the observed patient data, whether the patients that were clinically similar remain similar in the representation space, has not been systematically studied. This is the evaluation question that our correspondence score is designed to answer.

\subsection{Combining Intrinsic Measures into a Unified Score}

The intrinsic measures described in Section~2.1 are well established individually, but they are rarely combined in a principled way. Studies that report multiple intrinsic metrics typically present them side by side, leaving the reader to integrate disparate signals without guidance \cite{espadoto2019toward}. Visual tools such as Shepard diagrams and silhouette plots offer qualitative summaries but lack the quantitative precision needed for systematic model comparison \cite{lee2009quality, sedlmair2013taxonomy}. Interactive visualization systems enable exploration but do not produce reproducible, comparable scores \cite{chatterjee2021visual, wattenberg2016how}. Approaches that average metrics or assign fixed weights ignore the fact that different datasets have different structural properties, and that the relative importance of local versus global correspondence can vary considerably depending on the clinical context.

A further limitation of existing approaches is that they do not distinguish between different transformation directions. The structural demands of dimensionality reduction, where the goal is to retain meaningful relationships in a compressed space, differ from those of dimensionality expansion, where the concern is whether added dimensions carry new information or simply introduce redundancy. No existing approach applies direction-specific metric sets and adapts their combination to the characteristics of each dataset. These are precisely the gaps that our correspondence score addresses.

\subsection{Representations in Clinical Workflows}

Learned representations are not only used for prediction. In clinical research and practice, they support patient similarity search, unsupervised phenotype discovery, cohort stratification, and integration with other data modalities such as clinical notes or imaging \cite{Murali2023EHRKG, Huang2024HEART, Li2023HyperbolicPhenotypes, Pu2023GraphAD}. In each of these applications, the structure of the representation space directly determines which patients appear similar, how groups form, and which subgroups are identified. A representation that distorts patient neighborhoods will yield different, and potentially misleading, results in all of these analyses, even if it was validated and approved for a different predictive task.

This reuse pattern is expanding as foundation models enable representations to be shared across institutions and adapted to new purposes without retraining \cite{Timilsina2025HarmonizingFMs}. Yet no standard method exists to validate, before deployment, that a shared representation preserves the patient relationships present in the observed data. Our correspondence score provides exactly this: a reproducible instrument for assessing representation correspondence before representations are committed to clinical use.


\section{Methodology}
\label{sec:method}

This study evaluates how well different representation methods preserve the patient relationships present in observed clinical tabular data. To this end, we define a correspondence score and apply it systematically across multiple datasets, representation methods, and target dimensionalities. The evaluation covers both dimensionality reduction and expansion, reflecting the two directions in which representations are most commonly generated in biomedical informatics workflows. All experiments are conducted in an unsupervised, label-free setting: outcome labels were removed from all datasets prior to analysis, and no downstream task performance is used at any stage of the correspondence assessment.

\subsection{Datasets}
\label{sec:datasets}

We selected four publicly available clinical datasets, each built around routine blood count (CBC) measurements. CBC data are a particularly well-suited testbed for this study: they are standardized across institutions, widely collected as part of routine care, and their features encode clinically meaningful patient distinctions, including anemia subtypes, inflammatory profiles, and coagulation status. Using CBC-derived datasets allows correspondence to be assessed under a controlled feature schema while still capturing genuine clinical and demographic variability across patient populations.\newline
\textbf{CBC Panel (Kaggle) \cite{hematology_dataset}.}
A large CBC dataset (30,000 records, 24 features) comprising standard haematological indices including red cell parameters (haemoglobin, haematocrit, MCV, MCH, MCHC, RDW), white cell and differential counts, and platelet measurements. The dataset spans a clinically diverse population encompassing known haematological phenotypes such as iron-deficiency anaemia, macrocytic anaemia, and inflammatory leukocytosis, providing a richly structured testbed for assessing whether representations preserve meaningful patient distinctions. All features are numerical with no missing values.\newline
\textbf{Respiratory Infection Biomarkers (COVID-19) \cite{covid19_dataset}.}
A CBC cohort (377 records, 25 features) collected during COVID-19 clinical assessment, augmenting standard haematological indices with infection-related biomarkers including C-reactive protein, ferritin, lactate dehydrogenase, and D-dimer. Three categorical variables encode clinical status indicators. The combination of haematological and inflammatory markers introduces higher inter-feature correlation and greater clinical variability than a standard CBC panel, making this a more demanding test of local and global correspondence preservation.\newline
\textbf{Iraqi Hospital CBC Profiles \cite{iraq_dataset}.}
A CBC dataset (500 records, 21 features) collected from hospital records in Iraq, combining core haematological indices with demographic covariates including age and sex. Feature scales are heterogeneous across measurement types, and the population reflects the demographic and disease-burden characteristics of a Middle Eastern clinical setting. This dataset represents the kind of real-world distributional variability encountered when correspondence validation is applied outside standardised research cohorts.\newline
\textbf{UK Risk-Factor CBC Cohort \cite{liverpool_dataset}.}
A dataset (364 records, 12 features) linking routine CBC measurements to lifestyle and cardiovascular risk-factor variables collected in a UK population cohort. The feature set is smaller and more tightly scoped than the other datasets, combining continuous hematological measures with ancillary clinical context. This makes it a useful test of correspondence assessment under a compact, mixed feature schema.

\subsubsection{Harmonization and Preprocessing.}
To enable consistent comparison across datasets, we standardized feature names and measurement units (e.g., converting between g/dL and mmol/L where applicable), applied consistent encodings for binary categorical variables, and separated numerical from categorical fields for models that require typed inputs. All outcome labels present in the source datasets were removed before any analysis. Numerical features were normalized within each cohort independently. All neighborhood- and distance-based correspondence metrics were computed on the preprocessed feature space, using the same distance definition consistently when constructing patient relationships in the original and representation spaces.

\begin{table}[!t]
\caption{Summary statistics of the four clinical datasets used in this study, after harmonization and preprocessing.}
\renewcommand{\arraystretch}{1.3}
\setlength{\tabcolsep}{6pt}
\centering
\rowcolors{2}{gray!10}{white}
\resizebox{\textwidth}{!}{%
\begin{tabular}{lccccc}
\toprule
\textbf{Dataset} & \textbf{Samples} & \textbf{Features} & \textbf{Numeric} & \textbf{Categorical} & \textbf{Missing (\%)} \\
\midrule
CBC~\cite{hematology_dataset}      & 30{,}000 & 24 & 24 &  0 & 0.0 \\
COVID-19~\cite{covid19_dataset}    &      377 & 25 & 22 &  3 & 0.0 \\
Iraq~\cite{iraq_dataset}           &      500 & 21 & 21 &  0 & 0.0 \\
UK~\cite{liverpool_dataset}        &      364 & 12 & 12 &  0 & 0.0 \\
\bottomrule
\end{tabular}}
\label{tab:dataset_statistics}
\end{table}

\subsection{Representation Methods}
\label{sec:methods}

Representations were generated using two complementary families of methods: tabular foundation models and classical dimensionality transformation techniques. The two families are evaluated side by side to assess whether modern representation-learning architectures preserve patient relationships more faithfully than established alternatives.

\subsubsection{Tabular Foundation Models.}
Five foundation models were included, selected to represent the main architectural paradigms currently used for tabular representation learning.
\textit{TransTab} encodes both feature values and column names through a transformer, producing representations that can transfer across datasets with different feature schemas.
\textit{TabTransformer} applies self-attention to categorical features to capture contextual dependencies among table columns.
\textit{FT-Transformer} is a compact transformer encoder that processes numerical and categorical features jointly with minimal architectural overhead.
\textit{SAINT} combines row-wise and column-wise attention with contrastive pretraining, modeling both sample-level and feature-level structure simultaneously.
\textit{TabNet} uses a sequential attention mechanism to select relevant features at each step, producing sparse and interpretable representations with self-supervised training support.
Representations were extracted from the learned internal state of each model following training on each dataset.

\subsubsection{Classical Transformation Methods.}
Six classical methods were included as interpretable baselines, split by transformation direction.
For \textit{dimensionality reduction}: PCA (linear projection that maximizes variance), UMAP (nonlinear manifold-based projection), and VAE--Encoder (nonlinear compression through a learned probabilistic bottleneck).
For \textit{dimensionality expansion}: Random Projection (distance-preserving randomized linear expansion), Polynomial Feature Expansion (explicit generation of interaction and higher-order terms), and VAE--Decoder (nonlinear mapping to a higher-dimensional generative space).
The t-SNE method was excluded because it is designed for two- or three-dimensional visualization and does not generalize to the higher-dimensional representation sizes examined here.

Representations were generated at eight target dimensionalities: 8, 12, 16, 20, 32, 64, 96, and 128. For each dataset, dimensions below the original feature count constitute reduction; dimensions above constitute expansion. Foundation models are compared against reduction baselines (PCA, UMAP, VAE--Encoder) and expansion baselines (Random Projection, Polynomial Expansion, VAE--Decoder) separately, reflecting the direction of transformation each method performs.

\subsection{Correspondence Score}
\label{sec:score}

\begin{figure*}[!htbp]
\centering
\begin{subfigure}[t]{0.5\textwidth}
    \caption{Step 1: Computing direction-specific structure-aware metrics from the original and transformed data.}
    \centering
    \includegraphics[width=\linewidth]{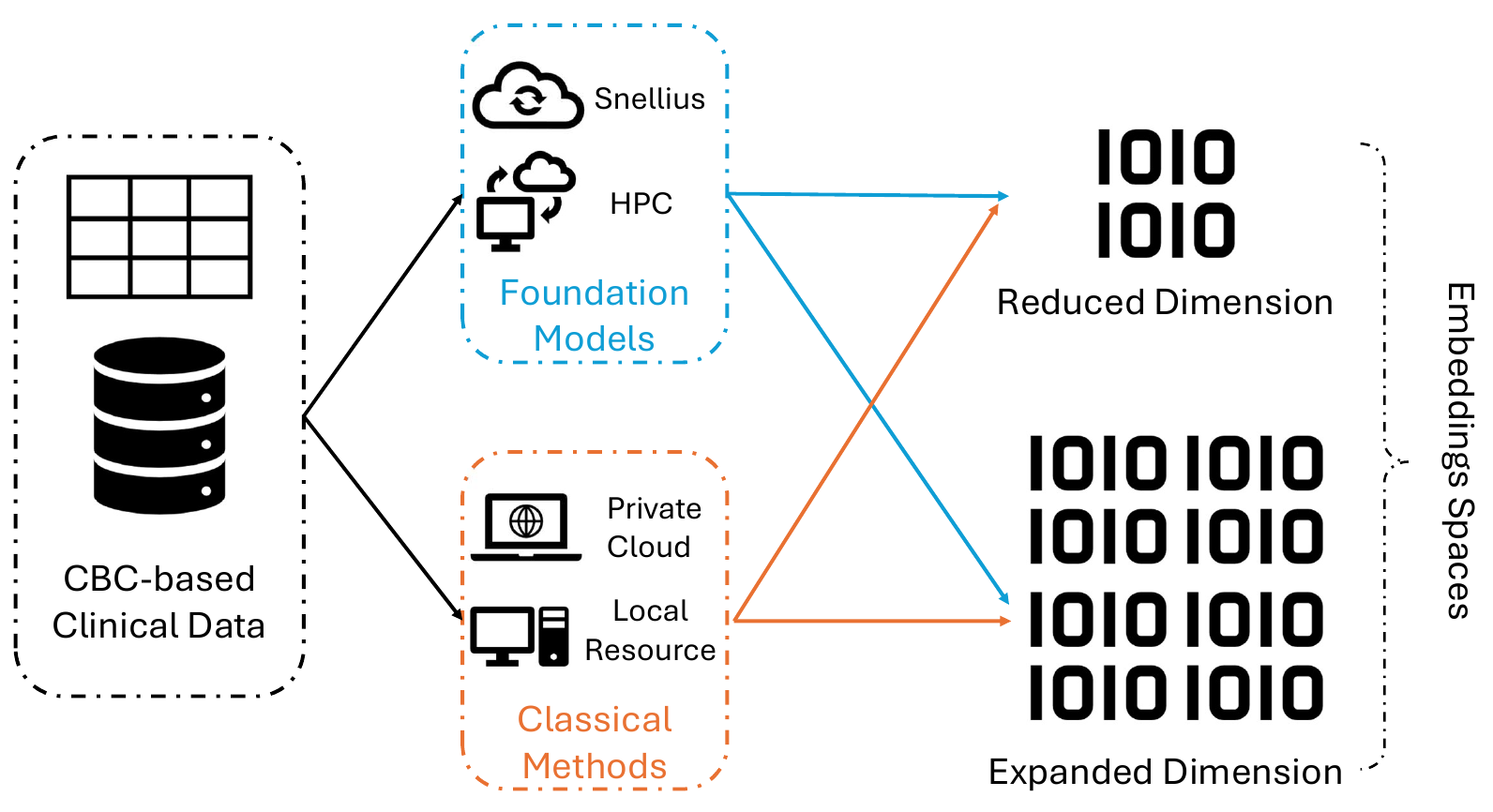}
    \label{fig:overview_a}
\end{subfigure}
\vspace{0cm}

\begin{subfigure}[t]{0.5\textwidth}
    \caption{Step 2: Estimating dataset-specific metric weights through Bayesian optimization.}
    \centering
    \includegraphics[width=\linewidth]{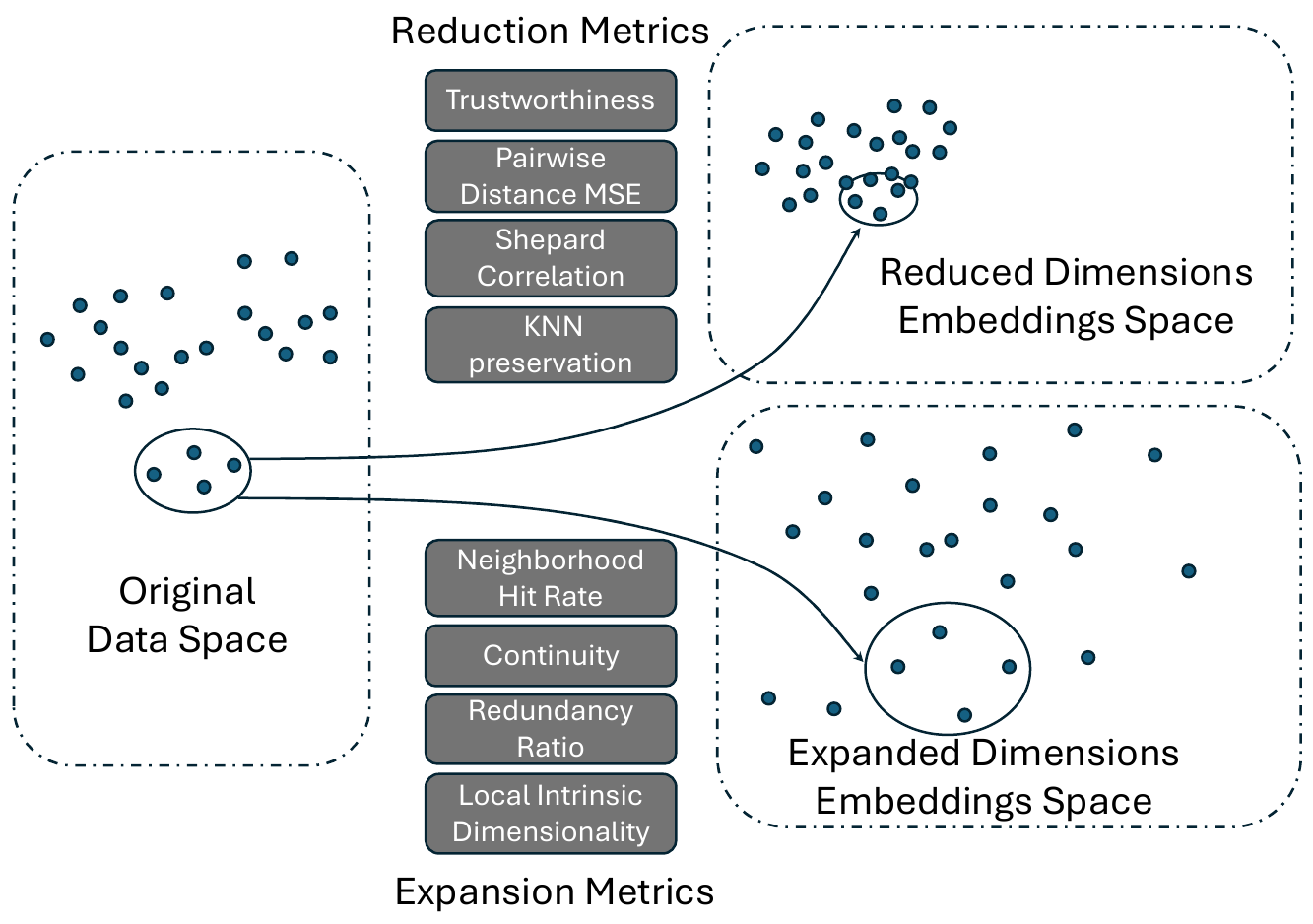}
    \label{fig:overview_b}
\end{subfigure}

\vspace{0cm}

\begin{subfigure}[t]{0.47\textwidth}
    \caption{Step 3: Deriving the CR-score and CE-score as direction-specific correspondence scores.}
    \centering
    \includegraphics[width=\linewidth]{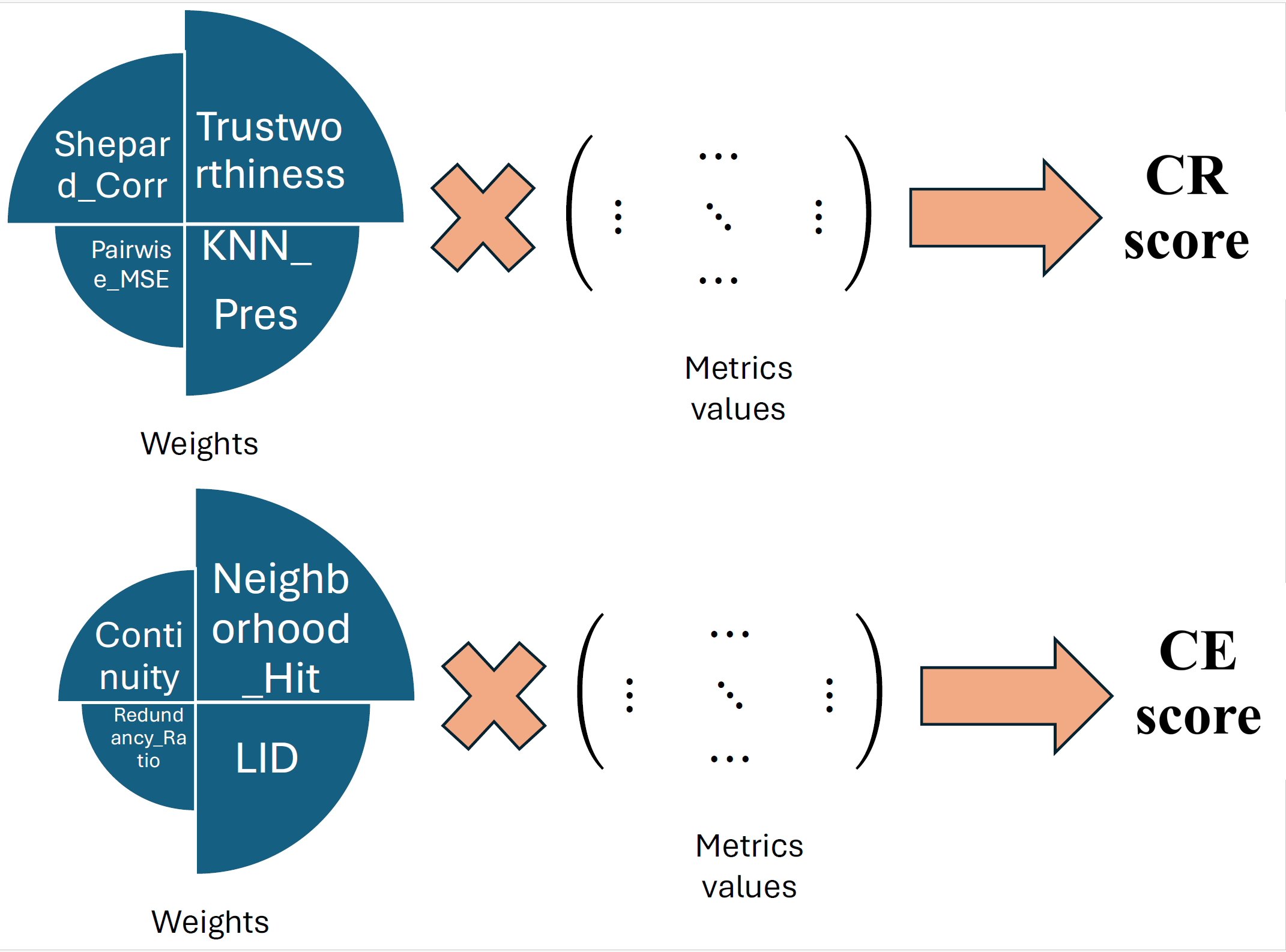}
    \label{fig:overview_d}
\end{subfigure}

\caption{Overview of the three-step correspondence score pipeline. Step 1 computes direction-specific structure-aware metrics for each representation. Step 2 estimates dataset-specific metric weights via Bayesian optimization. Step 3 derives the final CR-score and CE-score as weighted composites reflecting overall correspondence.}
\label{fig:framework_overview}
\end{figure*}

The correspondence score quantifies the extent to which patient relationships present in the observed data are preserved after transformation. It is computed in three steps: structural metric computation, metric aggregation via Bayesian optimization, and score derivation. The pipeline is illustrated in Figure~\ref{fig:framework_overview}.

\subsubsection{Step 1: Structural Metric Computation.}

The first step measures, for each generated representation, how well the patient relationships in the original data have been preserved. Because the structural demands of reduction and expansion differ, we use two separate sets of direction-specific metrics.

For \textit{dimensionality reduction}, the concern is whether compression retains meaningful neighborhood and distance structure. Four metrics are computed:
\textit{Trustworthiness} measures how many neighbors that appear close in the representation were genuinely close in the original data, penalizing the introduction of spurious neighbors.
\textit{Pairwise Distance MSE} measures the average squared difference between original and represented pairwise distances, capturing global distortion.
\textit{Shepard Correlation} measures the rank-order consistency of pairwise distances across the two spaces, reflecting whether the relative ordering of patient similarities is preserved.
\textit{kNN Preservation} measures what fraction of each patient's true nearest neighbors in the original data remain among their nearest neighbors after transformation.
Together, these four metrics assess both local correspondence (trustworthiness, kNN preservation) and global correspondence (distance MSE, Shepard correlation), two complementary aspects of what it means for a reduced representation to faithfully reflect the observed data.

For \textit{dimensionality expansion}, the concern shifts: when additional dimensions are added, do they carry new, meaningful structure or merely inflate the representation with redundancy? Four metrics address this:
\textit{Neighborhood Hit Rate} (NHR) measures the preservation of local neighborhood structure under expansion, assessing whether patients that are locally related in the original space remain locally related after additional dimensions are introduced.
\textit{Continuity} measures how many of each patient's original neighbors are recovered as neighbors in the expanded representation, penalizing the loss of genuine local structure.
\textit{Redundancy Ratio} measures the proportion of added dimension pairs whose absolute correlation exceeds 0.95, flagging dimensions that carry duplicate rather than new information.
\textit{Local Intrinsic Dimensionality} (LID) estimates the effective geometric complexity of local neighborhoods, indicating whether expansion genuinely enriches the representation or merely adds noise.
This set probes whether added capacity is informative: NHR and continuity assess whether patient neighborhoods survive, while redundancy ratio and LID assess whether the new dimensions contribute distinct structure.

\subsubsection{Step 2: Metric Aggregation via Bayesian Optimization.}

No single metric provides a complete picture of representation correspondence. Each captures one geometric property, and the relative importance of local versus global correspondence, or of neighborhood retention versus redundancy, varies across datasets depending on their structure, density, and feature composition. A fixed weighting scheme would impose the same priorities on all datasets regardless of these differences.

We therefore use Bayesian optimization to learn a dataset-specific weight vector over the direction-specific metrics, separately for reduction and expansion. Formally, the correspondence score is defined as:

\[
\mathcal{S} = \sum_{i=1}^{n} w_i \cdot M_i
\]

where $M_i$ is the normalized value of the $i$-th metric and $w_i$ is its weight, constrained so that $\sum w_i = 1$ and $0 \leq w_i \leq 1$. 
The task is to identify a weight vector $\mathbf{w}$ that provides a balanced composite of the normalized correspondence metrics for each dataset and transformation direction. The optimization objective maximizes the average composite correspondence score across the evaluated representations while penalizing highly concentrated weight distributions. This regularization prevents the composite score from being dominated by a single metric.

Because evaluating $\mathcal{S}$ for a candidate weight vector requires computing all metrics across all representations, the objective function is expensive and has no closed-form gradient. We approximate it with a Gaussian Process (GP) surrogate:

\[
f(\mathbf{w}) \sim \mathcal{GP}(\mu(\mathbf{w}),\, \sigma^2(\mathbf{w}))
\]

The surrogate models both the expected score and the uncertainty at any candidate $\mathbf{w}$. At each iteration, the Expected Improvement (EI) acquisition function selects the next candidate by balancing exploration of uncertain regions against exploitation of promising ones:

\[
\text{EI}(\mathbf{w}) = \left(\mu(\mathbf{w}) - \mathcal{S}^*\right)\Phi(Z) + \sigma(\mathbf{w})\,\phi(Z), \qquad Z = \frac{\mu(\mathbf{w}) - \mathcal{S}^*}{\sigma(\mathbf{w})}
\]

where $\mathcal{S}^*$ is the best score observed so far, and $\Phi$, $\phi$ are the standard normal CDF and PDF respectively. This process runs independently for each dataset and for each transformation direction, yielding eight weight vectors: one for dimensionality reduction and one for dimensionality expansion for each of the four datasets.

The resulting dataset-specific weights determine the relative contribution of each metric to the composite correspondence score while maintaining contributions from multiple aspects of representation quality. For example, if a dataset's reduced representations retain local neighborhoods well but distort global distances, Bayesian optimization will assign higher weights to trustworthiness and kNN preservation. Crucially, this procedure uses only structure-preservation metrics computed on fixed representations and does not rely on outcome labels.

It is worth noting that we do not introduce new metrics. The individual measures used here are well established in manifold learning and dimensionality reduction research. Our contribution is the principled organization of these metrics into direction-specific sets, and the data-adaptive aggregation that produces a single, interpretable correspondence score tailored to each dataset's structural characteristics.

\subsubsection{Step 3: Score Derivation.}

In the final step, the optimized weight vectors are applied to produce the direction-specific correspondence scores. Computing the weighted combination with the reduction metric set yields the \textit{Correspondence score for Reduction} (CR-score); computing it with the expansion metric set yields the \textit{Correspondence score for Expansion} (CE-score):

\[
\text{CR-score} = \sum_{i=1}^{n_r} w_i^{(r)} \cdot M_i^{(r)}, \qquad \text{CE-score} = \sum_{i=1}^{n_e} w_i^{(e)} \cdot M_i^{(e)}
\]

where superscripts $(r)$ and $(e)$ denote reduction and expansion respectively. This produces one score per model--dataset--dimension configuration, enabling examination of how correspondence changes as the target dimensionality increases or decreases.

To summarize performance across dimensionalities and enable model-level comparison, we compute the average CR-score over all reduction dimensions and the average CE-score over all expansion dimensions for each model--dataset pair. By construction, higher CR-score and CE-score values correspond to stronger preservation of patient relationships relative to the observed data.

\subsection{Implementation}

All experiments were implemented in Python 3.9. The foundation models (TransTab, SAINT, TabTransformer, FT-Transformer, TabNet) were implemented in PyTorch, with representations extracted from their learned internal states following training. Classical transformation methods were implemented using scikit-learn, including PCA, UMAP, random projection, polynomial feature expansion, and the encoder and decoder components of a $\beta$-VAE. Metric computations relied on scikit-learn, SciPy, and custom NumPy functions.

Computationally intensive foundation models were run on a hybrid setup of a local GPU workstation and the Snellius national supercomputing cluster. Metric evaluation and Bayesian optimization were performed on CPU. Bayesian optimization was implemented using scikit-optimize with Gaussian-process surrogates and Expected Improvement. All random seeds were fixed, preprocessing steps were standardized across datasets, and each model--dimension configuration was run through a unified pipeline to ensure reproducibility. The full codebase is publicly available.\footnote{https://github.com/majid-lotfian/Embedding-Space-Discovery}


\section{Results}
\label{sec:results}

We evaluate the correspondence score across four clinical datasets, five foundation models, and six classical transformation techniques, across eight target representation dimensionalities spanning dimensionality reduction and expansion, with the applicable direction determined by the original dimensionality of each dataset. Results are reported separately for dimensionality reduction and expansion, first examining the dataset-specific metric weights produced by Bayesian optimization, then tracing how correspondence changes with dimensionality, and finally comparing model families through their aggregated scores.

\subsection{Dataset-Specific Metric Weights Reflect Structural Diversity}

The Bayesian optimization procedure yielded distinct weight profiles for each dataset and each transformation direction (Tables~\ref{tab:opt_weights_reduction} and \ref{tab:opt_weights_expansion}). This variation is itself an informative result: it confirms that no single metric captures what matters most across all clinical datasets, and that a fixed weighting scheme would misrepresent correspondence in at least some settings.

\begin{table}[t]
\caption{Dataset-specific metric weights for computing the CR-score, derived through Bayesian optimization. Higher weights indicate greater structural importance for a given dataset in the dimensionality reduction setting.}
\label{tab:opt_weights_reduction}
\centering
\rowcolors{2}{gray!10}{white}
\resizebox{\textwidth}{!}{%
\begin{tabular}{lcccc}
\toprule
\textbf{Dataset} & \textbf{Trustworthiness} & \textbf{kNN Preservation} & \textbf{Pairwise Dist. MSE} & \textbf{Shepard Correlation} \\
\midrule
CBC~\cite{hematology_dataset}      & 0.382 & 0.248 & 0.144 & 0.226 \\
COVID-19~\cite{covid19_dataset}    & 0.583 & 0.073 & 0.070 & 0.274 \\
Iraq~\cite{iraq_dataset}           & 0.231 & 0.236 & 0.343 & 0.191 \\
UK~\cite{liverpool_dataset}        & 0.259 & 0.209 & 0.220 & 0.312 \\
\bottomrule
\end{tabular}}
\end{table}

\vspace{1em}

In the \textit{reduction} setting (Table~\ref{tab:opt_weights_reduction}), trustworthiness received the highest weight in the CBC and COVID-19 datasets, indicating that local neighborhood integrity was the most sensitive structural property under compression in those cohorts. The Iraq dataset, by contrast, placed more weight on pairwise distance MSE, suggesting that global distance fidelity was the dominant concern, likely reflecting greater feature heterogeneity and population diversity in that cohort. Shepard correlation was weighted highly in the UK and COVID-19 datasets, pointing to the importance of rank-order distance consistency in those cohorts. The kNN preservation metric contributed moderately and stably across all datasets, providing a reliable local signal without dominating the composite score in any single setting.

\begin{table}[t]
\caption{Dataset-specific metric weights for computing the CE-score, derived through Bayesian optimization. Higher weights indicate greater structural importance for a given dataset in the dimensionality expansion setting.}
\label{tab:opt_weights_expansion}
\centering
\rowcolors{2}{gray!10}{white}
\resizebox{\textwidth}{!}{%
\begin{tabular}{lcccc}
\toprule
\textbf{Dataset} & \textbf{Continuity} & \textbf{LID} & \textbf{Neighborhood Hit Rate} & \textbf{Redundancy Ratio} \\
\midrule
CBC~\cite{hematology_dataset}      & 0.127 & 0.121 & 0.737 & 0.016 \\
COVID-19~\cite{covid19_dataset}    & 0.498 & 0.142 & 0.285 & 0.075 \\
Iraq~\cite{iraq_dataset}           & 0.014 & 0.165 & 0.808 & 0.014 \\
UK~\cite{liverpool_dataset}        & 0.055 & 0.194 & 0.703 & 0.048 \\
\bottomrule
\end{tabular}}
\end{table}

In the \textit{expansion} setting (Table~\ref{tab:opt_weights_expansion}), the weight profiles shifted considerably. For the CBC, Iraq, and UK datasets, Neighborhood Hit Rate (NHR) emerged as the dominant metric, with weights of 0.737, 0.808, and 0.703 respectively, indicating that preservation of local neighborhood structure contributed strongly to the expansion correspondence score in these datasets. The COVID-19 dataset is the notable exception: here, Continuity received the highest weight (0.498), suggesting that the recovery of true neighbors in the expanded space mattered more than cluster membership retention, likely reflecting the richer local variability in that cohort. Weight distributions were also more concentrated in the expansion setting than in reduction, one metric typically dominated, reflecting that expansion quality is driven by a clearer primary concern in each dataset. Redundancy Ratio and Local Intrinsic Dimensionality consistently received low weights across all datasets, suggesting that these properties were either well controlled or dominated by the stronger neighborhood-based signal. The direction-specific weighting strategy is thus empirically justified: what matters for evaluating compression differs from what matters for evaluating enrichment, and both differ across clinical datasets.

\subsection{Correspondence Peaks at Intermediate Dimensionalities}

\begin{figure*}[!t]
    \begin{center}
        \subfloat[CBC \cite{hematology_dataset}]{%
            \includegraphics[clip,width=0.48\linewidth]{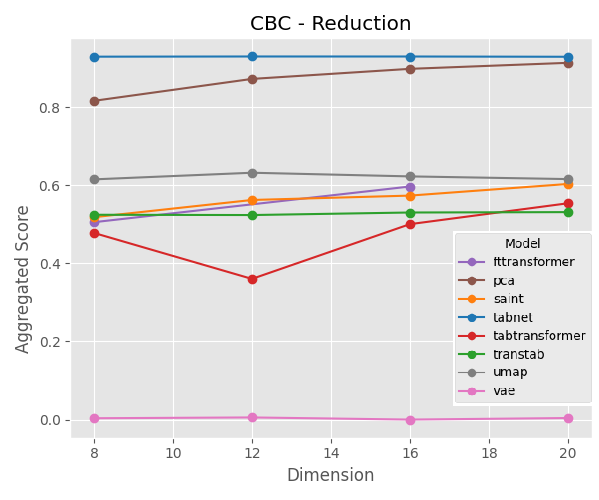}%
        }
        \subfloat[COVID-19 \cite{covid19_dataset}]{%
            \includegraphics[clip,width=0.48\linewidth]{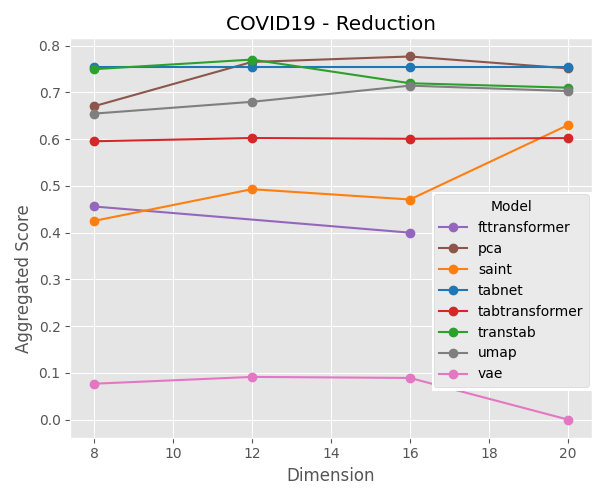}%
        }
        \vspace{1em}
        \subfloat[Iraq \cite{iraq_dataset}]{%
            \includegraphics[clip,width=0.48\linewidth]{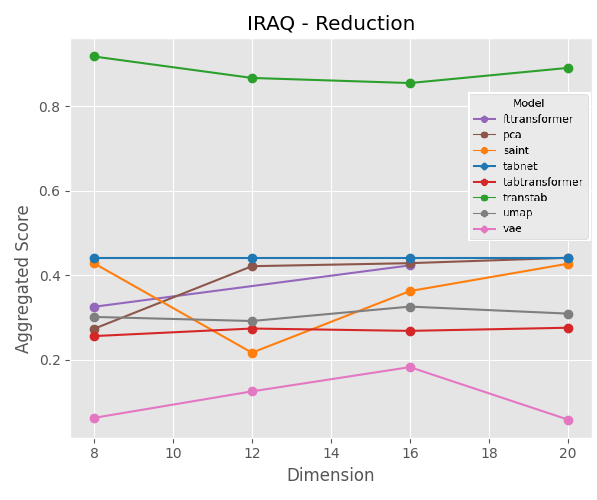}%
        }
        \subfloat[UK \cite{liverpool_dataset}]{%
            \includegraphics[clip,width=0.48\linewidth]{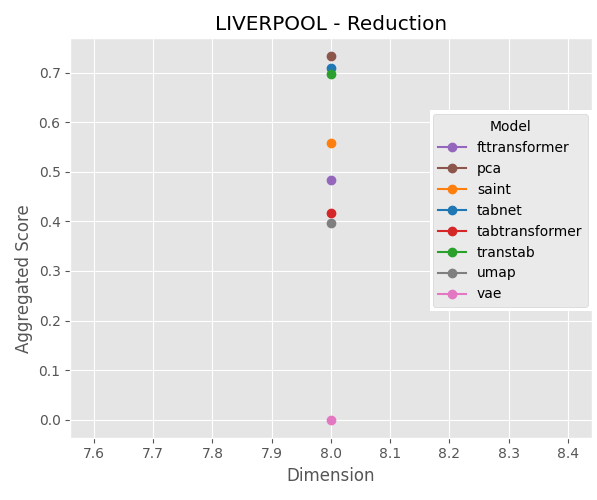}%
        }
        \caption{CR-score trends across representation dimensionalities for each dataset. Each curve traces how correspondence changes with target dimension for a given model in the reduction setting. \textcolor{mypurple}{$\bullet$} fttransformer, \textcolor{brown}{$\bullet$} pca, \textcolor{orange}{$\bullet$} saint, \textcolor{blue}{$\bullet$} tabnet, \textcolor{red}{$\bullet$} tabtransformer, \textcolor{mygreen}{$\bullet$} transtab, \textcolor{gray}{$\bullet$} umap, \textcolor{mypink}{$\bullet$} vae.}
        \label{fig:CR_trends}
    \end{center}
\end{figure*}

Figure~\ref{fig:CR_trends} shows the CR-score trends across all datasets in the reduction setting. 
Across the evaluated dimensionality ranges, correspondence generally showed its strongest values at intermediate dimensionalities, increased from the smallest sizes (8--12 dimensions), reached a peak in the moderate range (approximately 16--32 dimensions), and then plateaued or declined at larger sizes. This non-linear pattern suggests that moderate compression allows noise and redundancy in the original data to be filtered while preserving meaningful patient structure. Over-compression, forcing the data into too few dimensions, loses clinically relevant distinctions. But adding more dimensions beyond the structural content of the data does not help either; it introduces noise that dilutes rather than enriches correspondence.
This pattern was most apparent in datasets for which multiple reduction and expansion settings were available, while the UK dataset provided only a single reduction setting and therefore does not support the same comparison in the reduction direction. 

Classical methods such as PCA and UMAP perform competitively at low-to-moderate dimensionalities, where linear or manifold-based projections are sufficient to retain primary structure. At higher dimensions within the reduction range, foundation models tend to pull ahead, likely because their attention-based encoding can capture more complex, non-linear relationships among features.

\begin{figure*}[!t]
    \begin{center}
        \subfloat[CBC \cite{hematology_dataset}]{%
            \includegraphics[clip,width=0.48\textwidth]{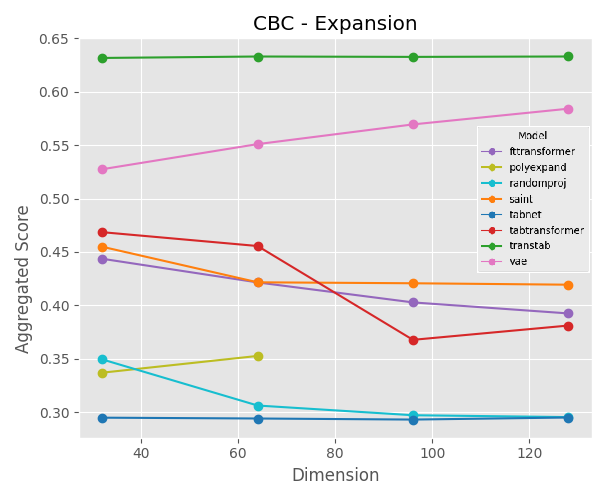}%
        }
        \subfloat[COVID-19 \cite{covid19_dataset}]{%
            \includegraphics[clip,width=0.48\textwidth]{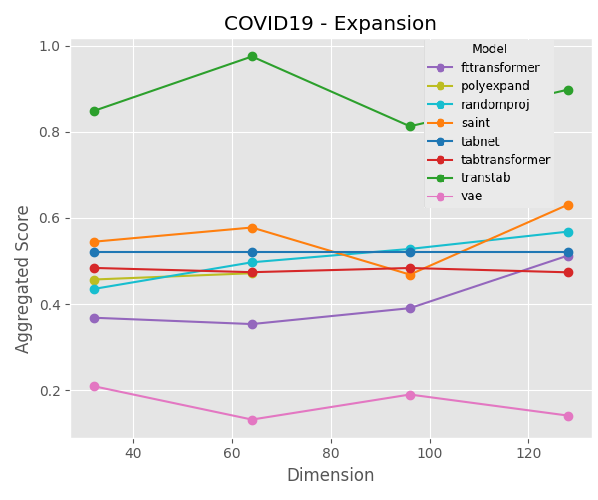}%
        }
        \vspace{1em}
        \subfloat[Iraq \cite{iraq_dataset}]{%
            \includegraphics[clip,width=0.48\textwidth]{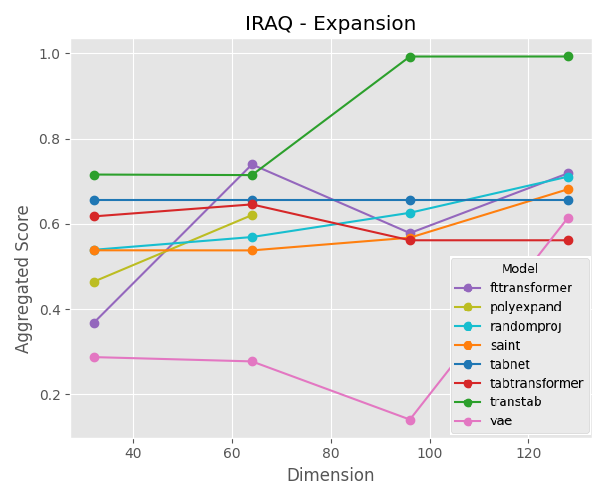}%
        }
        \subfloat[UK \cite{liverpool_dataset}]{%
            \includegraphics[clip,width=0.48\textwidth]{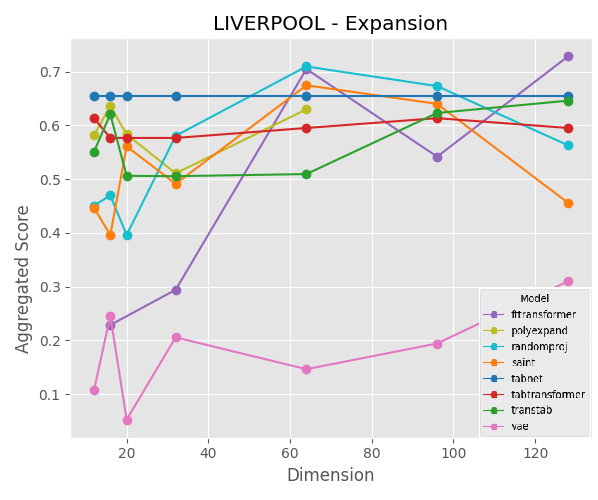}%
        }
        \caption{CE-score trends across representation dimensionalities for each dataset. Each curve traces how correspondence changes with target dimension for a given model in the expansion setting. \textcolor{mypurple}{$\bullet$} fttransformer, \textcolor{cedar}{$\bullet$} polynomial expansion, \textcolor{myblue}{$\bullet$} random projection,\textcolor{orange}{$\bullet$} saint, \textcolor{blue}{$\bullet$} tabnet, \textcolor{red}{$\bullet$} tabtransformer, \textcolor{mygreen}{$\bullet$} transtab,  \textcolor{mypink}{$\bullet$} vae.}
        \label{fig:CE_trends}
    \end{center}
\end{figure*}

In the expansion setting (Figure~\ref{fig:CE_trends}), CE-score trends are more dataset-dependent. For the CBC and UK datasets, correspondence improves steadily up to approximately 32--64 dimensions and then saturates, suggesting that these datasets have a limited amount of structure that additional dimensions can usefully encode. The COVID-19 dataset shows a different pattern: some models, particularly TabTransformer and TransTab, continue to gain correspondence at higher dimensions, likely because the richer feature content and clinical variability in that cohort benefit from the additional representational capacity.

Classical expansion methods, including polynomial feature expansion and random projection, show limited adaptability across the dimension range. They perform adequately at lower expansion sizes, but fail to improve correspondence as dimensions grow, and in some cases actively degrade it. This is consistent with the expectation that fixed transformation rules cannot adjust to the distributional properties of each dataset in the way that learned models can.

\subsection{Foundation Models Preserve Patient Relationships More Consistently}

\begin{figure}[t]
    \centering
    \includegraphics[width=\linewidth]{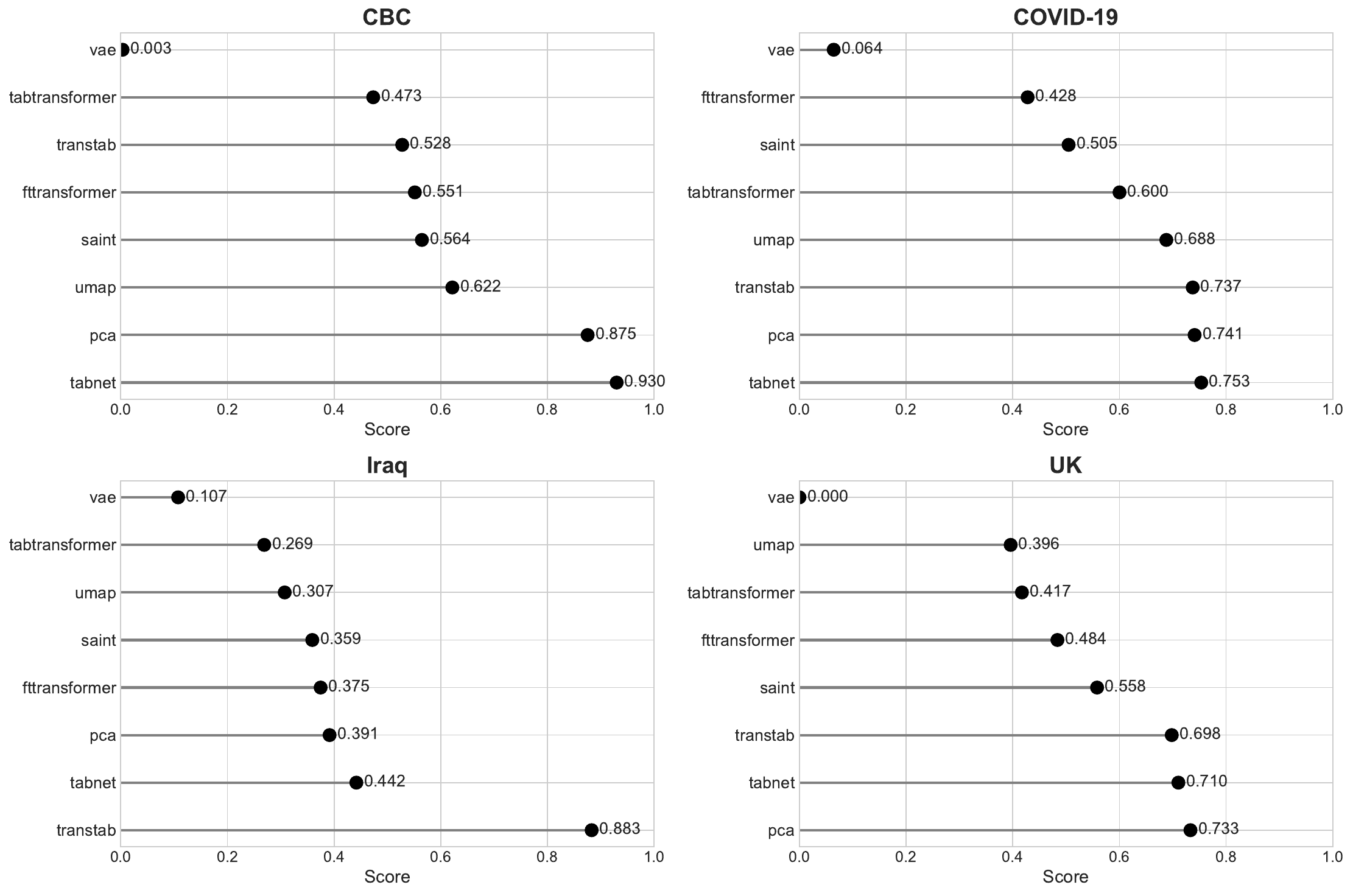}
    \caption{Average CR-scores per model and dataset, aggregated across all reduction dimensionalities.}
    \label{fig:CR_overal}
\end{figure}

\begin{figure}[t]
    \centering
    \includegraphics[width=\linewidth]{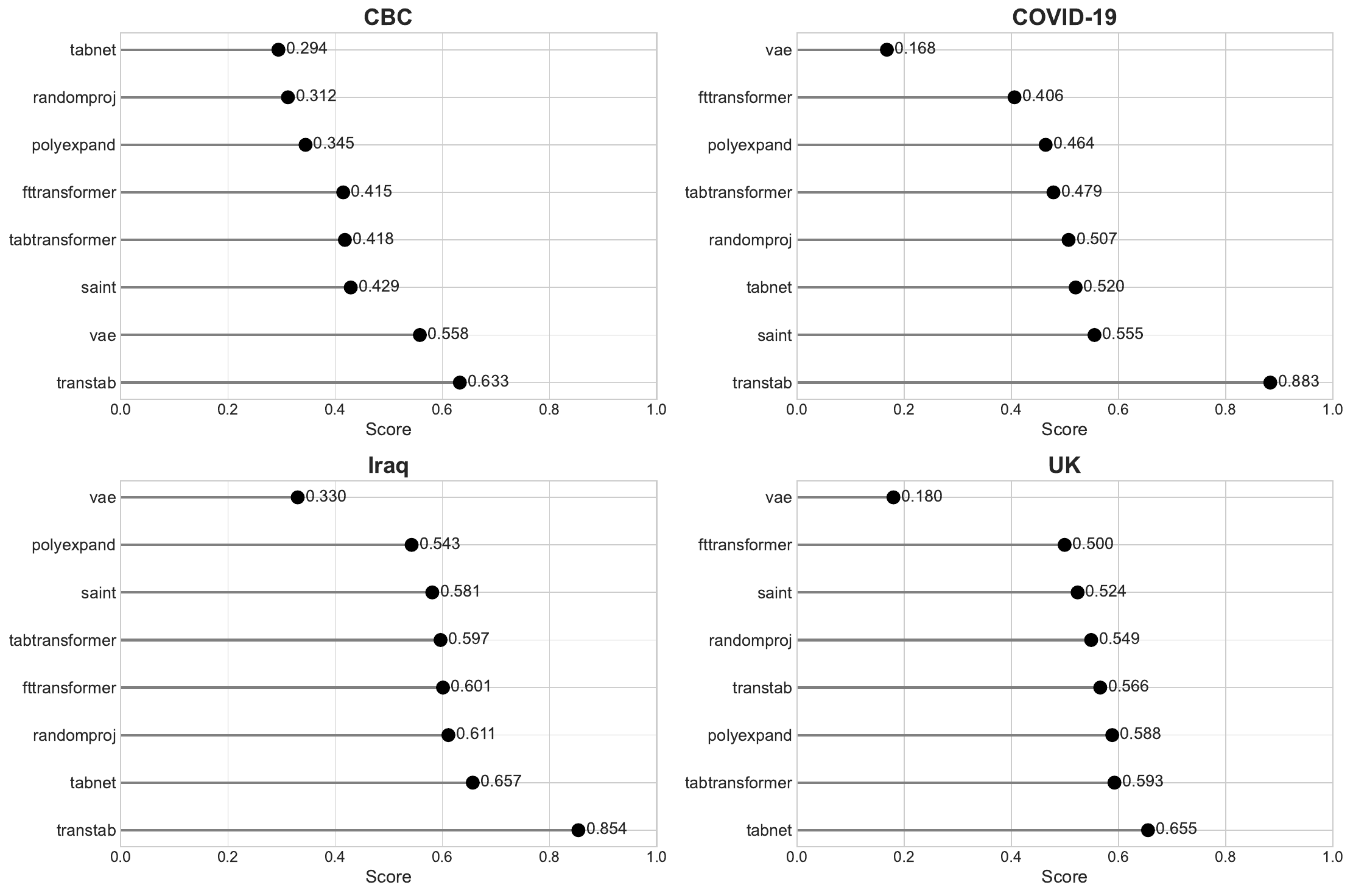}
    \caption{Average CE-scores per model and dataset, aggregated across all expansion dimensionalities.}
    \label{fig:CE_overal}
\end{figure}

Figures~\ref{fig:CR_overal} and \ref{fig:CE_overal} show the average CR-score and CE-score per model within each dataset, aggregated across all evaluated dimensionalities. These summaries reveal systematic differences between model families that are consistent across datasets and transformation directions.

In the \textit{reduction} setting (Figure~\ref{fig:CR_overal}), foundation models frequently outperform classical techniques. The differences are not trivial: TransTab exceeds PCA by over 125\% on the Iraq dataset, reflecting its ability to preserve complex patient relationships in a heterogeneous clinical cohort. To interpret this concretely: a 100\% improvement means that the baseline method preserved very little of the original patient structure, and the alternative recovered much of what was lost, for instance, keeping clinically distinct patient groups separated rather than merged. PCA retains an advantage on the more linearly structured CBC and UK datasets, outperforming TransTab by 39.7\% and 4.8\% respectively, which is consistent with PCA's strength on data where variance structure is the primary organizational principle. On the COVID-19 dataset, the two methods perform comparably. Even where TransTab does not lead, other foundation models contribute robustly: TabNet outperforms PCA by 5.5\% on the CBC dataset, and SAINT performs competitively across all four cohorts. VAE consistently underperforms in the reduction setting, which may reflect the inherent tension between its generative regularization objective and the geometric fidelity required for neighborhood preservation.

In the \textit{expansion} setting (Figure~\ref{fig:CE_overal}), the advantages of foundation models are even more pronounced. TransTab achieves the highest CE-score in three of the four datasets, including a 13.4\% improvement over the best classical method on the CBC dataset. On the COVID-19 and Iraq datasets, it maintains substantial leads over polynomial expansion and random projection. TabTransformer and FT-Transformer edge ahead in the UK dataset, but the margin is narrow. VAE continues to underperform across expansion tasks.

These results are notable for a specific reason: The evaluated models represent widely used deep tabular architectures and were trained within the experimental setting rather than being pretrained on large external tabular corpora. Future work should therefore examine whether the observed correspondence patterns generalize to large pretrained tabular foundation models, and yet they preserved patient relationships more faithfully than classical methods in most settings. This advantage may partly reflect architectural properties that allow these models to capture complex dependencies between features and preserve them in the learned representation. From a clinical AI development perspective, the implication is that foundation models represent a more reliable choice when the intended downstream use includes tasks sensitive to patient neighborhood structure, such as similarity search or cohort construction.

Because each dataset has distinct structural characteristics, in feature dimensionality, patient group separability, and distributional shape, we deliberately do not report a single aggregate score across datasets. Averaging across structurally heterogeneous cohorts would conflate different evaluation contexts and obscure the dataset-specific patterns that are themselves informative. Per-dataset scores are reported throughout to preserve interpretability.

\section{Discussion}
\label{sec:dis}

The results of this study support a clear and practically important message: task-based evaluation of learned clinical representations is insufficient on its own. Two representations of the same clinical dataset can achieve comparable predictive performance while differing substantially in how well they correspond to the patient relationships present in the observed data. These differences are not random, they are systematic functions of model architecture and representation dimensionality, and they cannot be detected without a validation step independent of downstream task performance. The correspondence score introduced in this study provides that step. We discuss the main findings in turn, with attention to their clinical implications.

\textbf{Correspondence captures what task performance cannot.}
The most fundamental result of this study is that representation correspondence and downstream task performance measure different things. A representation optimized to predict readmission will organize patients according to readmission-relevant patterns. It may achieve high accuracy, yet simultaneously merge patients with distinct anemia subtypes or separate patients with similar inflammatory profiles, differences that are invisible to the accuracy metric but clinically consequential when the same representation is reused for similarity search or phenotyping. The correspondence score makes this distinction visible. By measuring how well the structure of the observed data is preserved in the representation, it provides information that no extrinsic metric can supply. 
The results show that correspondence varies substantially, can differ by over 100\%, across representation models and dimensionalities, highlighting the importance of this distinction in any clinical setting where representations are reused.

\textbf{Correspondence depends on the dataset, not just the model.}
The Bayesian-optimized metric weights varied substantially across datasets and transformation directions, confirming that no single notion of correspondence applies universally to clinical tabular data. In some cohorts, the preservation of local patient neighborhoods was the dominant structural concern; in others, global distance relationships or dimensional redundancy mattered more. This context-dependence has a direct practical implication: correspondence validation should be performed on the specific dataset where a representation will be deployed, not assumed to transfer from benchmark evaluations. A model that performs well on one clinical dataset may preserve patient relationships poorly on another, even when the same measurement modality and feature schema are used. The dataset-adaptive weighting in our correspondence score makes this variation explicit rather than hiding it behind a fixed scoring convention.

\textbf{Representation dimensionality is a clinical design choice.}
A consistent finding across all datasets is that correspondence does not improve monotonically with representation size. Instead, it follows a non-linear trajectory: patient relationships are best preserved at intermediate dimensionalities, and both over-compression and excessive expansion degrade correspondence, though for different reasons. Over-compression forces multiple patient relationships into too few coordinates, merging distinct clinical groups; excessive expansion dilutes meaningful neighborhood structure with redundant or uninformative dimensions. The practical implication is that representation dimensionality should not be treated as a secondary hyperparameter tuned solely for downstream task performance. It is a design choice with direct consequences for the clinical analyses that depend on patient similarity. The CR-score and CE-score provide a concrete tool for identifying the dimensionality range where correspondence peaks, information that task-based evaluation cannot provide.

\textbf{Foundation models better preserve clinically meaningful patient structure.}
Across most datasets and transformation directions, tabular foundation models preserved patient relationships more faithfully than classical dimensionality transformation techniques. This held without pretraining on large tabular corpora, a distinction from the large language and vision models that typically benefit from scale. The advantage appears to stem from architectural properties: attention-based models can weigh feature interactions contextually, adapting their encoding to the specific structure of each patient record, rather than applying a fixed transformation rule. VAE-based representations consistently underperformed in both reduction and expansion settings, reflecting the known tension between the generative regularization that VAE training imposes and the geometric fidelity that correspondence requires. For clinical AI developers, these findings provide principled, pre-deployment guidance for model selection: when representations will be reused for similarity-sensitive tasks, attention-based foundation models are the more reliable choice, and this recommendation is now empirically grounded in correspondence rather than inferred from task performance.

\textbf{The correspondence score enables a new kind of clinical AI validation.}
In laboratory medicine, an assay is validated before it is used clinically, not by asking whether it correlates with a clinical outcome, but by verifying that it actually measures what it claims to measure. Learned representations are used as analytical instruments in an analogous way, yet they are rarely subjected to equivalent scrutiny before deployment. The correspondence score introduced here provides a direct parallel: it asks whether the representation, as an analytical instrument, faithfully reflects the patient relationships in the observed data before it is applied to downstream analyses. Applied as a routine pre-deployment validation step, it allows clinical AI developers to identify representations whose correspondence is inadequate before they influence similarity searches, cohort constructions, or phenotyping results. It also enables systematic comparison of model choices and dimensionality settings on a given clinical dataset, without requiring labeled outcomes, making it applicable precisely in the early-stage settings where labels are often scarce or not yet defined.

\textbf{Reusability and transferability of representations.}
A growing practice in clinical AI is to share representations across institutions or to reuse representations trained for one task in analyses designed for another. The correspondence score is directly relevant to both scenarios. A representation with high correspondence can be expected to preserve patient neighborhood structure regardless of the specific task it is applied to, because it reflects the observed data rather than a single task's optimization objective. Conversely, a representation with poor correspondence may perform well on its training task but fail silently when applied to a different clinical question. Integrating correspondence validation into the process of sharing or reusing representations, alongside conventional performance reporting, would provide a more complete picture of their fitness for clinical use.

\textbf{Generalizability and limitations.}
The empirical evaluation in this study focuses on routine blood count data, a well-standardized clinical modality available across institutions and with established clinical interpretation. This choice was deliberate: it provides a controlled and clinically grounded testbed where meaningful patient distinctions are known and where the effects of representation methods on correspondence can be studied under a consistent feature schema. The correspondence score itself is not specific to this modality; it operates on any tabular representation of patient data. Future work should extend the empirical validation to broader EHR-derived datasets, vital signs, chemistry panels, heterogeneous clinical records, to further establish the score's generalizability across clinical contexts. It will also be important to investigate how CR-score and CE-score values relate to downstream task performance across specific clinical applications, providing empirical evidence that stronger correspondence translates into more reliable clinical analyses. Extensions to longitudinal, graph-based, and multimodal clinical data, and examination of whether the metric-weighting strategy generalizes across datasets and institutions, represent further directions that would strengthen the role of correspondence validation as a standard step in clinical AI development.


\section{Conclusion}
\label{sec:con}

Learned representations of clinical data are increasingly reused across applications such as patient similarity search, cohort construction, phenotype discovery, and cross-institutional data analysis. These applications rely on the assumption that the representation adequately reflects the clinical structure present in the original data. Despite the growing use of learned representations in biomedical informatics, there is currently no standard approach for assessing this correspondence before representations are reused in downstream analyses.

This study introduces a correspondence score, comprising the score for Reduction (CR-score) and for Expansion (CE-score), to address this gap. The score combines complementary metrics measuring the preserved relationships through dataset-adaptive Bayesian optimization, producing a single interpretable measure of how well a learned representation corresponds to the original patient data. It requires no outcome labels, and can be applied to a wide range of tabular clinical datasets prior to downstream use.

Evaluation across four routine blood count datasets, five tabular foundation models, and six classical transformation techniques demonstrates that correspondence provides a complementary perspective on representations beyond conventional task-based evaluation. The correspondence score directly assesses how well relationships in the original data are preserved, providing information that is not reflected by predictive performance alone. Correspondence was highest for representations of intermediate dimensionality, while both lower- and higher-dimensional representations introduced measurable deviations from the original data structure, suggesting that representation dimensionality is an important design choice rather than merely a technical hyperparameter. In addition, tabular foundation models generally achieved higher and more consistent correspondence than classical transformation methods, supporting their use in applications where preservation of patient relationships and data structure is important.

Together, these findings position correspondence assessment as a practical validation step for representation reuse in clinical AI. Before a learned representation is used to identify similar patients, construct clinical cohorts, perform subgroup discovery, or support exploratory analyses, its correspondence with the original data can now be assessed directly, quantitatively, and without requiring labeled outcomes. Future work will investigate the relationship between correspondence and downstream performance across specific clinical applications, extend the framework to longitudinal and multimodal data, and examine the generalizability of the metric-weighting strategy across datasets and institutions. Incorporating correspondence validation into routine clinical AI workflows may ultimately support more reliable, interpretable, and clinically meaningful use of learned patient representations in biomedical research and practice.

\section*{Code and Data Availability}
\noindent
All code used for embedding generation, metric computation, and Bayesian
optimization is available at: \\ \url{https://github.com/majid-lotfian/Embedding-Space-Discovery}.\\
All datasets are publicly accessible under their respective licenses.

\begingroup
\scriptsize   
\bibliographystyle{unsrt}  
\bibliography{ref}
\endgroup

\end{document}